\documentclass[prb,amsmath,amssymb,superscriptaddress,twocolumn]{revtex4}
\usepackage{graphicx}
\usepackage{bbm}
\usepackage{bbold}
\usepackage{bm}
\newcommand{\al}{\alpha}

\newcommand{\ba}{{\bf a}}
\newcommand{\br}{{\bf r}}
\newcommand{\bx}{{\bf x}}
\newcommand{\by}{{\bf y}}
\newcommand{\bk}{{\bf k}}

\newcommand{\bp}{{\bf p}}

\newcommand{\bv}{{\bf v}}

\newcommand{\bz}{{\bf z}}
\newcommand{\bA}{{\bf A}}

\newcommand{\eps}{\epsilon}

\DeclareMathAlphabet{\mathpzc}{OT1}{pzc}{m}{it} 
\begin{document}
\title{Quantum oscillations of the specific heat in $d$-wave superconductors with loop current order}
\author{Luyang Wang}
\author{Oskar Vafek}
\affiliation{National High Magnetic Field Laboratory and Department
of Physics, Florida State University, Tallahassee, Florida 32306,
USA}
\address{}

\date{\today}
\begin{abstract}
We report numerical results of quantum oscillations of the specific
heat in the vortex state of a $d_{x^2-y^2}$-wave superconductor in
the presence of loop current order\cite{Varma97,Varma99}, which gives rise to Fermi
pockets coexisting with nodal $d_{x^2-y^2}$-wave superconductivity.
Within a lattice tight-binding model, we find that in an intermediate temperature range,
the oscillations seem to
approximately follow Onsager relation with an effective charge
comparable to the electric charge. However, the quasiparticle
spectrum does not resemble Landau levels. In order to understand
the origin of the oscillations, we also perform Franz-Tesanovic
transformation in the presence of the loop order and find that in
addition to scalar and Berry potentials\cite{FT}, one component of the gauge
invariant superfluid velocity couples to the low lying Dirac particles as a component of a vector
potential. The magnetic field
associated with this vector potential vanishes on average but is
highly non-uniform in the magnetic unit cell. We attribute the quantum oscillations to this field. We also compare the
results with the model without the loop order but with Zeeman-like
coupling which also induces Fermi pockets in the superconducting
state.
\end{abstract}
\maketitle

\section{Introduction}
\begin{figure}[t]
\begin{center}
\includegraphics[width=0.4\textwidth]{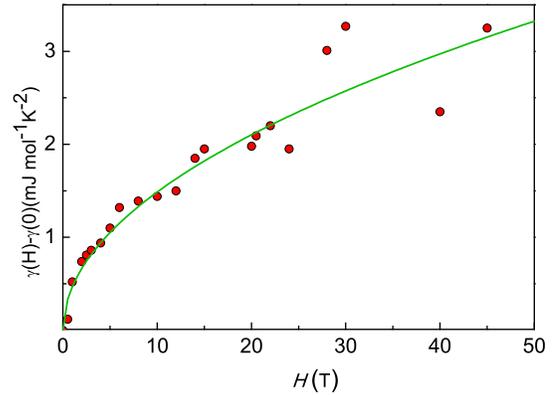}
\end{center}
\caption{$\gamma(H)-\gamma(0)$ for YB$_2$Cu$_3$O$_{6.56}$ (courtesy of S. Riggs). Discrete data points (red circles) are $C(T,H)/T$ extrapolated to $T=0$ in fixed magnetic fields, with $\gamma(0)$ subtracted. Green curve is $A_c\sqrt{H}$, where $A_c=0.47\rm{mJ/(mol\ K^2\ T^{1/2})}$.} \label{fig:experiment}
\end{figure}

Coexistence of $d$-wave superconductivity and Fermi pockets in
underdoped high temperature cuprate superconductors has been
suggested by recent quantum oscillation
experiments\cite{Doiron-Leyraud,Leboeuf,Sebastian,Yelland,Bangura,Audouard,Singleton,Riggs,Ramshaw}.
Whether the Fermi pockets are electron-like or
hole-like, and whether there is one or more than one pocket, is still under intense debate. In the
present work, we focus on quantum oscillations of the specific heat,
measurements of which have been
presented in Ref.\cite{Riggs}. The experimental data is shown in Fig.\ref{fig:experiment}. With the application of a magnetic
field $H$, the non-oscillatory component of the Sommerfeld coefficient $\gamma(H)$ of the ultrapure YB$_2$Cu$_3$O$_{6.56}$ exhibits
$\sqrt{H}$ behavior, which is consistent
with the $d$-wave superconductivity in the vortex state. Remarkably, this field dependence persists well into the resistive state. In addition, there are several
signatures of the existence of Fermi pockets. First, the zero field
Sommerfeld coefficient, $\gamma(0)\sim \rm{1.9mJ/(mol\ K^2)}$, is finite, indicating {\it finite}
density of states at zero energy in zero field. Note that the low
energy quasiparticles (QP's) of $d_{x^2-y^2}$-wave superconductors are characterized by the
linear Dirac-like dispersion near four nodal points. This results in
linearly {\it vanishing} density of states at zero energy. Although the finite density of states may in principle be induced by impurity
disorder, the YBCO samples under study\cite{Riggs} are believed to be too pure to account for the measured value of $\gamma(0)$. The high purity is consistent with the observation of the quantum oscillations as well as the extracted values of the Dingle temperature. Therefore, the physical origin of the nonzero $\gamma(0)$ is most likely intrinsic to this system. Second, the oscillatory component of $\gamma(H)$
exhibits quantum oscillations in high magnetic fields, periodic in $1/H$, which can be well fitted\cite{Doiron-Leyraud,Leboeuf,Sebastian,Yelland,Bangura,Audouard,Singleton,Riggs,Ramshaw} by Lifshitz-Kosevich (LK) formula.
Such phenomenology is quite remarkable. On the one hand, the quantum oscillations appear to be due to Landau level quantization of the electron orbits,
and indicate the existence of Fermi pockets, while on the other, $\gamma(H)\sim\sqrt{H}$ is a signature of $d_{x^2-y^2}$-wave superconducting gap and the vortex state.

The origin of Fermi pockets in the superconducting state has been
under debate. One possible scenario is that the Fermi pockets arise
from the one-dimensional CuO chains hybridized into the BaO layers\cite{Riggs}. If not gapped by a
proximity effect down to the lowest temperatures ($\sim$1K) at which $\gamma(0)$ was extracted, such Fermi pockets could result in a finite $\gamma(0)$, as well as quantum oscillations of $\gamma(H)$. This would account for the main experimental features.

In this paper we critically examine another scenario in which loop
current order induces Fermi pockets in the $d_{x^2-y^2}$ superconductor\cite{BergKivelson,KivelsonVarma}. In such an ordered phase,
charge currents circulate within each unit cell (as shown in
Fig.\ref{fig:disp}), breaking time reversal symmetry and inversion
symmetry, but not their product or the discrete translational symmetry of the lattice. Two Fermi pockets of Bogoliubov quasiparticles, one
electron-like and one hole-like, are formed, giving rise to
nonzero density of states at zero energy \cite{BergKivelson}. This
may account for the finite $\gamma(0)$ and the question is whether it can
also cause quantum oscillations of $\gamma(H)$ in high fields. In a
superconductor, the Bogoliubov QP's are linear combinations of
electrons and holes, and therefore do not carry definite charge. On the other hand, the QP's in Fermi liquids do carry
definite charge\cite{A&M}. Therefore, it is not {\it a priori} obvious whether
there are any quantum oscillations at all, and if yes, whether the
oscillations obey Onsager relation as in Fermi liquids. In this
paper, we show that the effective magnetic field experienced by the
Dirac quasiparticles in the loop order state vanishes on average and does not lead to Landau
quantization. As such any oscillations do not follow the detailed
LK phenomenology. Nevertheless, the effective magnetic field
experienced by the Dirac quasiparticles is highly non-uniform and
in an intermediate temperature range the quantum oscillations of
the specific heat appear to approximately obey Onsager relation,
with an effective charge comparable to the electric charge.

We investigate the oscillations of the specific heat in both the
tight-binding lattice formulation and in the continuum formulation. We
assume that the vortices form a square Abrikosov lattice. In the
tight-binding lattice formulation, we take the vortices to sit
inside the plaquettes of the two-dimensional CuO$_2$-like plane. In
each magnetic unit cell, there are two singly quantized vortices
with flux $hc/2e$. Since the vortices are placed at the centers of
the plaquettes, the vortex lattice has to be commensurate with the
underlying tight-binding lattice. This prevents us from sweeping the
magnetic field continuously. Instead, in this case, we sweep the (Bogoliubov)
Fermi pocket area by varying the overall magnitude of the loop
current order in fixed magnetic fields, and investigate the
dependence of the density of states and the specific heat on the Fermi pocket area. The results for the density of states are shown in Fig.\ref{fig:DOS}, where we also show that they clearly differ from the density of states of Landau quantized anisotropic Dirac fermions. Nevertheless, as shown in Figs.(\ref{fig:4H_2K}-\ref{fig:Vortex position}), we find
that in an intermediate, magnetic field dependent, temperature window, the specific heat
exhibits oscillations as a function of Fermi pocket area for the four values of the magnetic field studied, ranging from 7.7T to 35.6T. At the same time, the non-oscillatory component of $\gamma(H)$ does not follow $\sqrt{H}$ behavior (see Fig.\ref{fig:4H_5K}).

To further understand the origin of this effect, we complement the
tight-binding calculations with an approximate continuum
formulation. To this end, we linearize the Hamiltonian in the
vicinity of the four nodal points, perform the
Franz-Tesanovic transformation\cite{FT}, calculate the quasiparticle
spectrum numerically using plane-wave diagonalization and calculate
the specific heat. While we are well aware of the subtleties with the
large gauge invariance\cite{Vafek2001}
we are merely interested in the overall qualitative aspects of the results
and their dependence on the strength of loop order and
magnetic field. We find
that the result obtained using this second method is consistent with
the one obtained in the tight-binding lattice formulation. The second
method offers an additional advantage in that the external magnetic
field can be changed continuously; the resulting oscillations of
specific heat are shown in Figs.(\ref{fig:40vs28linH}-\ref{fig:Cv_H}).

Finally, we compare these results with the results obtained by varying the Zeeman energy but without loop
current order. The Zeeman term shifts all four nodal points, resulting in four Fermi pockets. In this case, the oscillations do not obey Onsager relation at all.

Our paper is organized as follows. In Sec.II, we set up both the lattice and the continuum Hamiltonians, and calculate the zero field spectrum. In Sec.III, we present the numerical results for the quantum oscillations of specific heat as a function of the loop current order and Zeeman energy, and the density of states in the lattice formulation. In Sec.IV, we present the numerical results for the oscillations as a function of the loop current order and the magnetic field in the linearized problem. In Sec.V, we discuss our results.

\section{Formalism: BdG Hamiltonian and Singular Gauge Transformation with Loop Current Order}
\subsection{Lattice formulation}
\begin{figure}[t]
\begin{center}
\includegraphics[width=0.2\textwidth]{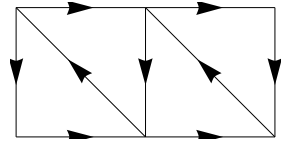}
\end{center}
\begin{center}
\includegraphics[width=0.4\textwidth]{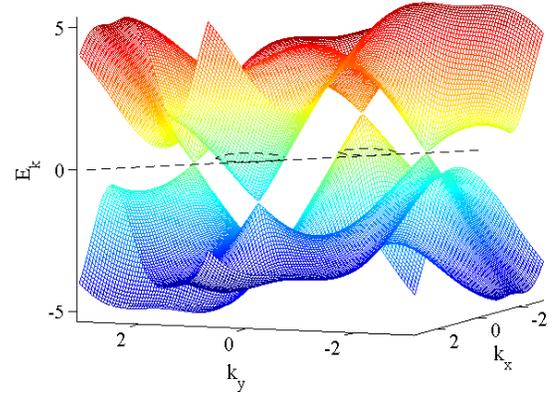}
\end{center}
\caption{(Upper) The connectivity of the loop current network. (Lower) The dispersion of $d$-wave superconductor with loop current order in the first Brillouin zone. For clarity, the anisotropy is set to 1, and the strength of loop current order $J$=0.5. The dashed line indicates the zero energy.} \label{fig:disp}
\end{figure}

We model the CuO$_2$ plane in YBCO as a tight-binding lattice with lattice constant $a$, which may be set to 1 for convenience. (When converting to real units, we use $a=0.38$nm.) When an external magnetic field $H$ in the range $H_{c1}<H<H_{c2}$ is applied, the $d$-wave superconductor enters vortex state and the vortices form an Abrikosov lattice. We assume that a square vortex lattice is formed with magnetic unit cell $\ell_B\times\ell_B$, where the magnetic length $\ell_B$ is defined through the flux quantum $\phi_0=hc/e$ as $\ell_B=\sqrt{\phi_0/H}$. In each magnetic unit cell, there are two singly quantized vortices, each of which carrying flux $hc/2e$. Our starting point is the Hamiltonian with nearest neighbor hopping, $d$-wave pairing and loop current order on the underlying tight-binding lattice in the presence of a magnetic field,
\begin{eqnarray}
\mathcal{H}=\mathcal{H}_0+\mathcal{H}_J
\end{eqnarray}
where\cite{VafekMelikyan}
\begin{eqnarray}\label{eq:H0}
\mathcal{H}_0&=&-t\sum_{\langle\br\br'\rangle\sigma}\left(e^{-iA_{\br\br'}}c^\dagger_{\br\sigma}c_{\br'\sigma}+h.c.\right)\nonumber \\
&&+\sum_{\langle\br\br'\rangle}\left(\Delta_{\br\br'}(c^{\dagger}_{\br\uparrow}c^{\dagger}_{\br'\downarrow} +c^{\dagger}_{\br'\uparrow}c^{\dagger}_{\br\downarrow})+h.c.\right)
\end{eqnarray}
and the Hamiltonian for loop current order is\cite{BergKivelson}
\begin{eqnarray}\label{eq:HJ}
\mathcal{H}_J=\sum_{\br\br'\sigma}(-iJ_{\br\br'}c^{\dagger}_{\br\sigma}c_{\br'\sigma}+h.c.).
\end{eqnarray}
In Eq.(\ref{eq:H0}), the sums are over nearest neighbors $\langle\br\br'\rangle$, and $\sigma$ denotes the spin. In the symmetric gauge, the magnetic flux $\Phi$ through an elementary plaquette enters the Peierls factor via $\bA_{\br\br+\hat{\bx}}=-\pi y\Phi/\phi_0$ and $\bA_{\br\br+\hat{\by}}=\pi x\Phi/\phi_0$. The $d$-wave pairing field in the vortex lattice is $\Delta_{\br\br'}=\eta_{\br-\br'}\Delta_0 e^{i\theta_{\br\br'}}$, where $\eta_\delta=+(-)$ if $\delta||\hat{\bx}(\hat{\by})$, and the Ansatz for the pair phases is\cite{VafekMelikyan}
\begin{eqnarray}
e^{i\theta_{\br\br'}}\equiv \frac{e^{i\phi(\br)}+e^{i\phi(\br')}}{|e^{i\phi(\br)}+e^{i\phi(\br')}|},
\end{eqnarray}
where $\nabla\times\nabla\phi(\br)=2\pi\hat{\bz}\sum_{i}\delta(\br-\br_i)$ and $\nabla\cdot\nabla\phi(\br)=0$ where $\br_i$ denotes the vortex positions. In Eq.(\ref{eq:HJ}), the connectivity of the loop current network $J_{\br\br'}$ is determined according to Fig.\ref{fig:disp}, and in zero field all the nonzero currents have the same magnitude $J$\cite{Varma97,Varma99,BergKivelson}. In a finite magnetic field, we have, explicitly,
\begin{eqnarray}
\mathcal{H}_J&=&-iJ\sum_{\br}\left(e^{-iA_{\br\br+\hat{\bx}}}c^{\dagger}_{\br\sigma}c_{\br+\hat{\bx}\sigma} +e^{-iA_{\br+\hat{\by}\br}}c^{\dagger}_{\br+\hat{\by}\sigma}c_{\br\sigma}\right.\nonumber\\
&&\left.+e^{-iA_{\br+\hat{\bx},\br+\hat{\by}}}c^{\dagger}_{\br+\hat{\bx}\sigma}c_{\br+\hat{\by}\sigma}\right)+h.c.
\end{eqnarray}
where $A_{\br+\hat{\bx},\br+\hat{\by}}=-\pi(x+y+1)\Phi/\phi_0$.
We perform the particle-hole transformation,
\begin{eqnarray}
c^{\dagger}_{\br\uparrow}&=&d^{\dagger}_{\br\uparrow},\\
c^{\dagger}_{\br\downarrow}&=&d_{\br\downarrow}.
\end{eqnarray}
Then the diagonalization of the Hamiltonian is equivalent to the solution of the Bogoliubov-de Gennes (BdG) equation $\hat{\mathcal{H}}\psi_\br=E\psi_\br$ where the lattice operator
\begin{eqnarray}
\mathcal{\hat{\mathcal{H}}}=\left(\begin{array}{cc}\mathcal{\hat{E}}_\br+\mathcal{\hat{J}}_\br-\mu&\hat{\Delta}_\br\\ \hat{\Delta}_\br^*&-\mathcal{\hat{E}}^*_\br+\mathcal{\hat{J}}_\br^*+\mu\end{array}\right).
\end{eqnarray}
with $\mu$ being the chemical potential. The BdG Hamiltonian acts on
the two component Nambu spinor $\psi_\br=[u_\br,v_\br]^T$, and
$\mathcal{\hat{E}}$, $\mathcal{\hat{J}}$ and $\hat{\Delta}$ are
defined through their action on a lattice function $f_\br$ as
\begin{eqnarray}
\mathcal{\hat{E}}_\br f_\br&=&-t\sum_{\delta=\pm\hat{\bx},\pm\hat{\by}}e^{-i\bA_{\br\br+\delta}}f_{\br+\delta},\\
\mathcal{\hat{J}}_\br f_\br&=&-iJ\sum_{\delta=\hat{\bx},-\hat{\by},\hat{\by}-\hat{\bx}}e^{-i\bA_{\br\br+\delta}}f_{\br+\delta},\\
\hat{\Delta}_\br f_\br&=&\Delta_0\sum_{\delta=\pm\hat{\bx},\pm\hat{\by}}e^{i\theta_{\br\br+\delta}}\eta_\delta f_{\br+\delta},
\end{eqnarray}
\begin{figure}[t]
\begin{center}
\includegraphics[width=0.3\textwidth]{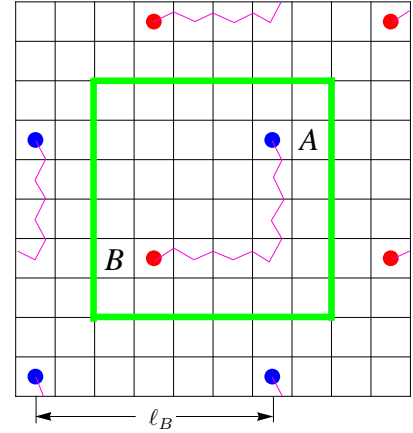}
\end{center}
\caption{Magnetic unit cell $\ell_B\times\ell_B$ containing $A$ and $B$ vortex joined by a branch-cut with $\ell_B=6a$.} \label{fig:MUC}
\end{figure}

The Hamiltonian is invariant under discrete translations followed by
a gauge transformation (magnetic translations). As shown in
Ref.\cite{FT}, it can be transformed into a periodic Hamiltonian by
a singular gauge transformation
\begin{eqnarray}\label{eq:FT}
\mathcal{U}=\left(\begin{array}{cc}e^{i\phi_e(\br)}&0\\0&e^{-i\phi_h(\br)}\end{array}\right)
\end{eqnarray}
where $\phi_e(\br)$ and $\phi_h(\br)$ satisfy
$\phi_e(\br)+\phi_h(\br)=\phi(\br)$. The vortices are divided into
two groups $A$ and $B$, and each magnetic unit cell contains one $A$
and one $B$ vortex, as shown in Fig.\ref{fig:MUC}. Then two phase
fields $\phi_A(\br)$ and $\phi_B(\br)$ are identified with
$\phi_e(\br)$ and $\phi_h(\br)$, respectively. If we choose
$\phi_A(\br)=\phi_B(\br)=\phi(\br)/2$, then the transformation
becomes $\mathcal{U}=\exp{\frac{i}{2}\sigma_3\phi(\br)}$. Connecting
pairs of vortices in one magnetic unit cell by a branch cut as shown
in Fig.\ref{fig:MUC}, we have
\begin{eqnarray}
e^{i\theta_{\br\br'}}e^{-\frac{i}{2}\phi(\br)}e^{-\frac{i}{2}\phi(\br')}=z_{2,\br\br'},
\end{eqnarray}
where as discussed in detail in Ref.\cite{VafekMelikyan}, the Z$_2$
field $z_{2,\br\br'}=1$ on each bond except the ones crossing the
branch cut where $z_{2,\br\br'}=-1$. Then the transformed
Hamiltonian
$\tilde{\mathcal{H}}=\mathcal{U}^{-1}\mathcal{\hat{H}}\mathcal{U}$
is
\begin{eqnarray}
\tilde{\mathcal{H}}=\sigma_3(\tilde{\mathcal{E}}_\br-\mu)+\sigma_1\tilde{\Delta}_\br+\tilde{\mathcal{J}}_\br\mathbb{1},
\end{eqnarray}
where the transformed lattice operators satisfy
\begin{eqnarray}
\tilde{\mathcal{E}}_\br\psi_\br&=&-t\sum_{\delta=\pm\hat{\bx},\pm\hat{\by}}z_{2,\br\br+\delta}\times e^{i\sigma_3 V_{\br\br+\delta}}\psi_{\br+\delta},\\
\tilde{\mathcal{J}}_\br\psi_\br&=&-iJ\sum_{\delta=\hat{\bx},-\hat{\by},\hat{\by}-\hat{\bx}}z_{2,\br\br+\delta}\times e^{i\sigma_3 V_{\br\br+\delta}}\psi_{\br+\delta},\\
\tilde{\Delta}_\br\psi_\br&=&\Delta_0\sum_{\delta=\pm\hat{\bx},\pm\hat{\by}}z_{2,\br\br+\delta}\times \eta_\delta\psi_{\br+\delta},
\end{eqnarray}
$\sigma$'s are Pauli matrices and $\mathbb{1}$ is the
identity matrix, and
\begin{eqnarray}
e^{iV_{\br\br'}}=\frac{1+e^{i(\phi(\br)-\phi(\br'))}}{|1+e^{i(\phi(\br)-\phi(\br'))}|}e^{-i\bA_{\br\br'}}.
\end{eqnarray}
The resulting Hamiltonian is invariant under magnetic translations
by $\ell_B$ in both directions, so it can be diagonalized in the
Bloch basis. The transformed Hamiltonian
$\mathcal{H}(\bk)=e^{-i\bk\br}\tilde{\mathcal{H}}e^{i\bk\br}$
becomes
\begin{eqnarray}\label{eq:BdG}
\mathcal{H}(\bk)=\sigma_3(\tilde{\mathcal{E}}_\br(\bk)-\mu)+\sigma_1\tilde{\Delta}_\br(\bk)+\tilde{\mathcal{J}}_\br(\bk)\mathbb{1}
\end{eqnarray}
where
\begin{eqnarray}
\tilde{\mathcal{E}}_\br(\bk)\psi_\br&=&-t\sum_{\delta=\pm\hat{\bx},\pm\hat{\by}}z_{2,\br\br+\delta}\times e^{i\sigma_3 V_{\br\br+\delta}}e^{i\bk\cdot\delta}\psi_{\br+\delta},\\
\tilde{\mathcal{J}}_\br(\bk)\psi_\br&=&-iJ\sum_{\delta=\hat{\bx},-\hat{\by},\hat{\by}-\hat{\bx}}z_{2,\br\br+\delta}\times e^{i\sigma_3 V_{\br\br+\delta}}e^{i\bk\cdot\delta}\psi_{\br+\delta},\\
\tilde{\Delta}_\br(\bk)\psi_\br&=&\Delta_0\sum_{\delta=\pm\hat{\bx},\pm\hat{\by}}z_{2,\br\br+\delta}\times \eta_\delta e^{i\bk\cdot\delta}\psi_{\br+\delta}.
\end{eqnarray}

\subsection{Zero-field spectrum}
In the absence of a magnetic field, the phase factors $e^{iA_{\br\br'}}$ and $e^{iV_{\br\br'}}$ become 1, and the Hamiltonian can be easily diagonalized, with eigenenergies
\begin{eqnarray}\label{eq:disp}
E_{\bk}=\pm\sqrt{\xi_\bk^2+\Delta_\bk^2}+2J[\sin k_x-\sin k_y+\sin(k_y-k_x)]
\end{eqnarray}
where $\xi_\bk=2t(\cos k_x+\cos k_y)-\mu$ and $\Delta_\bk=2\Delta_0(\cos k_x-\cos k_y)$. In the case with $J=0$, the four nodes of the spectrum are located at $(\pm k_D,\pm k_D)$ where
\begin{eqnarray}
k_D=\arccos{(\frac{\mu}{4t})}.
\end{eqnarray}
In the vicinity of each node, the dispersion can be linearized
\begin{eqnarray}
E_{\bk}=\sqrt{v_F^2\delta k_\perp^2+v_\Delta^2\delta k_\parallel^2},
\end{eqnarray}
where $\delta k_\perp(\delta k_\parallel)$ is the displacement of
the momentum from a node in the direction perpendicular (parallel)
to the Fermi surface, and the velocities are
\begin{eqnarray}\label{eq:vel}
v_F&=&2\sqrt{2}\sqrt{1-(\frac{\mu}{4})^2}t,\\
v_\Delta&=&2\sqrt{2}\sqrt{1-(\frac{\mu}{4})^2}\Delta_0,
\end{eqnarray}
where $\mu$'s are in units of $t$.
In the case $J\neq0$, the last term in Eq.(\ref{eq:disp}) near the $\pm (k_D,k_D)$-nodes is expanded as
\begin{eqnarray}
\frac{\sqrt{2}}{2}(4-\mu)J\delta k_\parallel
\end{eqnarray}
and near the $\pm (k_D,-k_D)$-nodes
\begin{eqnarray}\label{eq:J0}
\pm J_0-v_J\delta k_\perp,
\end{eqnarray}
where $J_0=J(1-\frac{\mu}{4})\sqrt{16-\mu^2}$ is the energy shift of
the nodes, and $v_J=\frac{\sqrt{2}}{4}J(-\mu^2
+2\mu+8)$. As a result of the shift, two Fermi pockets are induced,
as shown in Fig.\ref{fig:disp}, giving a finite density of states at
zero energy.

\subsection{Continuum formulation and the linearized Hamiltonian}
In the low temperature specific heat measurement, only the low
energy excitations contribute to the result. QP's near the
$\pm(k_D,k_D)$ nodes may be expected to result in the $\sqrt{H}$ behavior of the Sommerfeld
coefficient $\gamma(H)$. On the other hand, the low energy QP's near the Fermi surfaces at $\pm(k_D,-k_D)$ may be expected to give rise to the finite zero field
Sommerfeld coefficient $\gamma(0)$ and perhaps even the quantum oscillations in
high fields. To test this, we
formulate the continuum version of the BdG Hamiltonian in the
presence of loop current order, and linearize it near the four nodal points.

In the absence of loop current order, the continuum Hamiltonian
reads\cite{Vafek2001}
\begin{eqnarray}
\mathcal{H}=\left(\begin{array}{cc}\hat{\mathcal{H}}_e&\hat{\Delta}\\\hat{\Delta}^*&-\hat{\mathcal{H}}^*_e\end{array}\right),
\end{eqnarray}
with $\hat{\mathcal{H}}_e=1/2m(\hat{\bp}-e/c\bA)^2-\mu$,
$\hat{\bp}=-i\hbar\nabla$ the momentum operator, and $\nabla\times\bA=H\bf{z}$. In the following,
we choose the $x$-axis along the $(k_D,k_D)$ nodes and the $y$-axis along
the $(-k_D,k_D)$ nodes.
Then the gauge invariant $d$-wave pairing operator has the form
\begin{eqnarray}
\hat{\Delta}=\frac{1}{p_F^2}\{\hat{p}_x,\{\hat{p}_y,\Delta(\br)\}\}+\frac{i}{4p_F^2}\Delta(\br)(\hat{p}_x\hat{p}_y\phi),
\end{eqnarray}
with $p_F$ the Fermi momentum and $\phi$ the phase of the
superconducting gap $\Delta(\br)$. The curly bracket represents
symmetrization, $\{a,b\}=1/2(ab+ba)$. After the singular gauge
transformation (\ref{eq:FT}), the Hamiltonian becomes\cite{FT,Vafek2001}
\begin{eqnarray}
\left(\begin{array}{cc}\frac{1}{2m}(\hat{\bp}+m\bv_s^A)^2-\mu&\hat{D}\\\hat{D}&-\frac{1}{2m}(\hat{\bp}-m\bv_s^B)^2+\mu \end{array}\right),
\end{eqnarray}
where
$\hat{D}=\Delta_0/2p_F^2[\hat{p}_x+a_x][\hat{p}_y+a_y]+(x\leftrightarrow
y)$ and $\bv_s^\mu=1/m(\hbar\nabla\phi_\mu-e/c\bA)$ for $\mu=A,B$. The Berry vector potential\cite{FT}
$\ba=m/2(\bv_s^A-\bv_s^B)=\hbar/2(\nabla\phi_A-\nabla\phi_B)$.
The linearized approximation in the vicinity of one of the
$\pm(k_D,-k_D)$ nodes results in
\begin{equation}
\mathcal{H}_{N}=\mathcal{H}_0+\mathcal{H}',
\end{equation}
where
\begin{eqnarray}
\mathcal{H}_0=\left(\begin{array}{cc}v_F\hat{p}_y&v_\Delta\hat{p}_x\\v_\Delta\hat{p}_x&-v_F\hat{p}_y\end{array}\right)
\end{eqnarray}
is the free Dirac Hamiltonian and
\begin{eqnarray}
\mathcal{H}'=\left(\begin{array}{cc}mv_Fv_{sy}^A&v_\Delta a_x\\v_\Delta a_x&mv_F v_{sy}^B\end{array}\right).
\end{eqnarray}
In the above $v_F$ is the Fermi velocity and $v_\Delta=\Delta_0/p_F$ is the slope of
the gap at the node.
$\mathcal{H}_N$ can be written as
\begin{eqnarray}
\mathcal{H}_N=v_F(\hat{p}_y+a_y)\sigma_3+v_\Delta(\hat{p}_x+a_x)\sigma_1+mv_Fv_{sy},
\end{eqnarray}
where $\bv_s=(\bv_s^A+\bv_s^B)/2=1/m(\hbar/2\nabla\phi-e/c\bA)$ is the
superfluid velocity. From $\mathcal{H}_N$ it is readily seen that
$\ba$ couples to the Dirac fermions as a vector potential while
$\bv_s$ results in a Doppler shift. The magnetic field produced by
$\ba$ consists of a set of $\pm\pi$-flux delta function spikes at
the vortex cores and vanishes on average. It does not lead to Landau
level quantization\cite{FT}.

It is expected that physical quantities should be independent of the
choice of $A$ and $B$ sublattices, since there should be no physical
distinction between $A$ and $B$ vortices.
However, as discussed in Ref.\cite{Vafek2001}, two distinct
choices of $A$-$B$ sublattices as illustrated in Fig.2 of
Ref.\cite{Vafek2001} result in qualitatively similar but still
somewhat different band structures and densities of
states, particularly at higher energies. Despite significant
effort\cite{MelikyanTesanovic} this problem remains a bit of a
challenge: while the large gauge invariance is easily restored by
judicious enforcement of boundary conditions at vortex locations,
the interference among the nodes in a perfect vortex lattice obscures
the ultimate choice for these boundary conditions\cite{MelikyanTesanovic}.
At any rate, these mathematical subtleties are inherent only to the
linearized BdG Hamiltonian and they do not arise at all in the
tight-binding lattice formulation.

The linearized Hamiltonian associated with loop current
order can be derived from Eq.(\ref{eq:BdG}). Near one of the Fermi
pockets, it reads
\begin{eqnarray}
\mathcal{H}_J&=&(-v_J\hat{p}_y+J_0)\mathbb{1}.
\end{eqnarray}
In a magnetic field, after the singular gauge transformation, it
becomes
\begin{eqnarray}
-v_J(\hat{p}_y+a_y)\mathbb{1}-mv_Jv_{sy}\sigma_3+J_0\mathbb{1}.
\end{eqnarray}
Therefore, the full linearized Hamiltonian near one of the Fermi pockets is
\begin{eqnarray}\label{eq:linH}
\mathcal{H}&=&v_F(\hat{p}_y+a_y)\sigma_3+v_\Delta(\hat{p}_x+a_x)\sigma_1+mv_F v_{sy}\nonumber\\
&-&v_J(\hat{p}_y+a_y)-mv_Jv_{sy}\sigma_3+J_0,
\end{eqnarray}
from which it is seen that the superfluid velocity couples to the Bogoliubov QP's, in part, as a vector potential through loop current order,
\begin{eqnarray}
(v_F\hat{p}_y-mv_Jv_{sy})\sigma_3=v_F(\hat{p}_y-\frac{v_J}{v_F}(\frac{\hbar}{2}\partial_y\phi-\frac{e}{c}A_y))\sigma_3.
\end{eqnarray}
The effective vector potential $\bf{a_{eff}}$ has a zero
$x$-component, while the $y$-component is $(v_J/v_F)(c/e)mv_{sy}$.
The associated effective field, $\bf{b}_{eff}=\nabla\times
\ba_{eff}$, vanishes on average in the magnetic unit cell. The linearized Hamiltonian
near the $\pm(k_D,k_D)$-nodes, which are not shifted by the loop order, resembles Eq.(\ref{eq:linH}), but with $J_0=0$.

\section{Numerical Results}
At low temperature, the non-oscillatory part of the specific heat $C(T,H)$ is linear in the temperature $T$, for a Fermi liquid composed of Schr\"{o}dinger particles with $\bp^2/2m$ dispersion. The Sommerfeld coefficient can be defined as
$\gamma(H)=C(T,H)/T$. Experimentally, it has been found that the non-oscillatory part of
$\gamma(H)$ goes as $\sqrt{H}$ in low field, which is consistent
with the $d$-wave vortex state scenario. In the high field
$\gamma(H)$ also develops an oscillatory component, which obeys LK
formula\cite{Shoenberg}
\begin{eqnarray}\label{eq:LK}
&&C_{osc}(T,H)=\nonumber\\
&&-AT\sum_{p=1}^\infty R_D J_0(4\pi
p\frac{t_w}{\hbar\omega_c})\cos(2\pi
p(\frac{\mu}{\hbar\omega_c}-\frac{1}{2}))f''(x)\nonumber\\
\end{eqnarray}
where $A$ is a constant, $R_D=\exp{((-2\pi^2
pk_BT_D)/(\hbar\omega_c))}$ is the Dingle factor,
$\omega_c=eH/(m^*c)$ is the cyclotron frequency with $m^*$ the effective mass, $x=2\pi^2
pk_BT/(\hbar\omega_c)$, $f''(x)=x((1+\cosh^2 x)/\sinh^3 x-2\cosh
x/\sinh^2 x)$, $J_0$ is the Bessel function of the first kind (not to be confused with
the energy shift by loop current order), and $t_w$ is the $c$-axis
hopping energy. In the experiments only the first harmonic with
$p=1$ is identified. In the presence of loop current order, $J\neq0$, and the Sommerfeld coefficient becomes $\gamma(H,J)$. We will compare our results with the LK formula and show
that in the vortex state with loop current order the formula does
not hold.

In the appendix, we derive the formula for the oscillatory part of the specific heat assuming that Dirac particles with velocities $v_F$ and $v_\Delta$ couple minimally to the vector potential corresponding to a uniform magnetic field. As we stressed before, the $d$-wave Dirac particles do not have such coupling. Nevertheless, we find it useful to contrast our numerical finding to this analytical formula. In this case, the expression for $C_{osc}$ is similar to Eq.(\ref{eq:LK}), but there is an important difference. The effective mass in the above formula is replaced by $E_F/(v_Fv_\Delta)$. As such, the amplitude of the oscillations also depends on the Fermi energy in addition to the temperature and the magnetic field.

We use realistic values of physical quantities of YBCO as parameters
in our Hamiltonian. The Fermi velocity is taken to be
$v_F=2.15\times10^5$m/s, the lattice constant $a=0.38$nm, the doping
$15\%$, and the Dirac cone anisotropy $\al=14$. We first get
$\mu=0.297t$ from the doping, and then derive the nearest neighbor
hopping energy $t=0.132eV$ using Eq.(\ref{eq:vel}). Within our
method, we are not able to sweep the magnetic field continuously as
mentioned in the introduction. Instead, we sweep the Fermi pocket area by
varying the loop current order or Zeeman energy in a fixed magnetic
field. From this viewpoint, Onsager relation reads\cite{A&M}
\begin{eqnarray}\label{eq:Onsager}
A(\xi_{\nu+1})-A(\xi_\nu)=A_0
\end{eqnarray}
where $\xi_\nu$ is $\nu$th energy level when a magnetic field is
applied. This means that the period of oscillations, which is the
difference between the areas enclosed by the orbits of adjacent
energy levels in $k$-space, equals the area of the magnetic
Brillouin zone $A_0\equiv\frac{2\pi eH}{\hbar c}=\frac{4\pi^2
H}{\phi_0}=(\frac{2\pi}{\ell_B})^2$. We study the specific heat in
four different magnetic fields, with magnetic length $\ell_B=$
$60a$, $40a$, $36a$ and $28a$. For $a=0.38$nm this corresponds to field strengths 7.7T,
17.4T, 21.5T and 35.6T, respectively. Being fully aware of the caveat that for Dirac particles the {\it amplitude} of the oscillations of the specific heat may also depend on the Fermi pocket area, and therefore strictly speaking does not follow the Onsager relation, we investigate whether such relation holds in the $d$-wave superconducting state with loop current order.

We use Arnoldi algorithm to diagonalize the Hamiltonian (\ref{eq:BdG}). Only the
low energy bands need to be taken into account since the high energy bands give negligible contribution to the low temperature specific heat. Using $t=
1580$K, all bands below 100K are considered. This gives us enough
accuracy to determine the specific heat up to $\sim$10K. We find that a $40\times40$ mesh in
the first magnetic Brillouin zone (corresponding to a system with
$40\times40$ magnetic unit cells) gives convergent
results, showing little difference from that with a $80\times80$
mesh at the temperatures under study.

In what follows, we present results for the Sommerfeld
coefficient with loop current order $\gamma(H,J)$ in fixed magnetic
fields $H$ while $J$ is continuously swept, which are later compared with
the results from sweeping the Zeeman energy.

\subsection{Oscillations as a function of loop current order}
\subsubsection{Frequency of oscillations}
\begin{figure}[t]
\begin{center}
\includegraphics[width=0.4\textwidth]{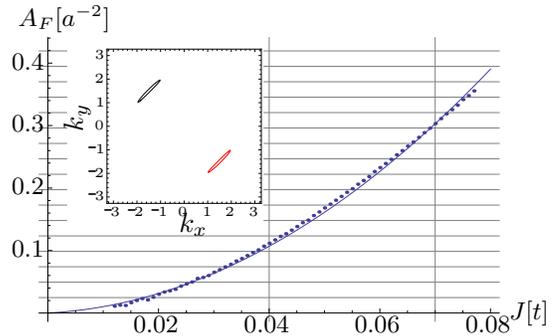}
\end{center}
\caption{The dependence of Fermi pocket area $A_F$ on loop current order $J$ in the $d$-wave superconducting state. The dots calculated from numerics are fitted by a parabola. $A_F$ is in unit of $a^{-2}$, where $a=0.38$nm is the linear size of the unit cell, and $J$ is in unit of the nearest neighbor hopping energy $t$. The anisotropy $\al=14$. The interval between adjacent horizontal grid lines is $A_0=(2\pi/\ell_B)^2$, where the magnetic length $\ell_B=40a$ in this figure. There are about 8 intervals between $J=0.04t$ and $0.07t$, marked by the vertical lines. (Inset) The electron-like and hole-like Fermi pockets in the 1st Brillouin zone at $J=0.05t$.} \label{fig:FSarea}
\end{figure}

In the presence of loop current order $J$, two Fermi pockets appear
in the $(k_D,-k_D)$ direction of the Brillouin
zone\cite{BergKivelson}, as shown in Fig.\ref{fig:disp}. At small
$J$, the area of each Fermi pocket $A_F$ is quadratic in $J$ since
the dispersion has a Dirac cone structure. Fig.\ref{fig:FSarea}
shows numerically calculated $A_F$ vs. $J$ for the tight binding model. The interval between two
adjacent horizontal lines is $A_0\equiv(2\pi/\ell_B)^2$, with the
magnetic length $\ell_B=40a$. We vary the Fermi pocket area by choosing
$J=t\sqrt{0.04^2+0.5\times 10^{-4}n}$, with the integer $n$ ranging
from $1$ to $70$. Then $J$ ranges approximately from $0.04t$ to
$0.07t$, within which there are about $8$ intervals of $A_0$.

\begin{figure}[t]
\begin{center}
\includegraphics[width=0.4\textwidth]{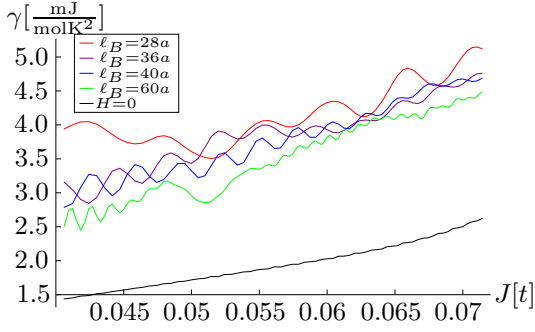}
\end{center}
 \caption{The Sommerfeld coefficient $\gamma(H,J)$ as a function of loop current order $J$ in a $d$-wave superconductor, in four different finite fields with magnetic length $\ell_B=28a$ (red), $36a$ (purple), $40a$ (blue) and $60a$ (green) and zero field (black). All curves are at 2K except that the $\ell_B=60a$ curve is at 1K.} \label{fig:4H_2K}
\end{figure}

\begin{figure}[t]
\begin{center}
\includegraphics[width=0.4\textwidth]{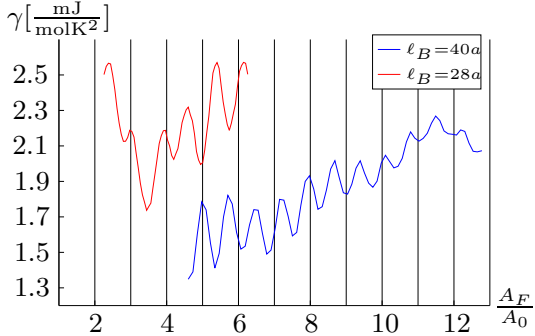}
\end{center}
 \caption{The Sommerfeld coefficient $\gamma(H,J)$ as a function of the rescaled Fermi pocket area $A_F/A_0$, for $\ell_B=40a$ (blue) and $28a$ (red) at 2K after the zero field background is subtracted.}\label{fig:40vs28}
\end{figure}

Fig.\ref{fig:4H_2K} shows $\gamma(H,J)$ vs. $J$ at low temperature
for the four different finite fields and zero field. All the curves are at 2K except the one
in the lowest field (green) which is at 1K. The zero field $\gamma(0,J)$ is
calculated using the dispersion Eq.(\ref{eq:disp}). For this temperature, the frequency of the
oscillations basically obeys Onsager relation Eq.(\ref{eq:Onsager}),
i.e. the frequency (or the number of periods) of oscillations is
proportional to the inverse of the magnetic field. A comparison of the oscillations at two
different fields as a function of the rescaled Fermi pocket area is
shown in Fig.\ref{fig:40vs28}, where the zero field background has
been subtracted. There are about 1.4 periods between two adjacent
vertical lines for both fields, suggesting that the Onsager relation
holds approximately, albeit with an effective charge $e^*\approx 0.7e$.

\begin{figure}[t]
\begin{center}
\includegraphics[width=0.4\textwidth]{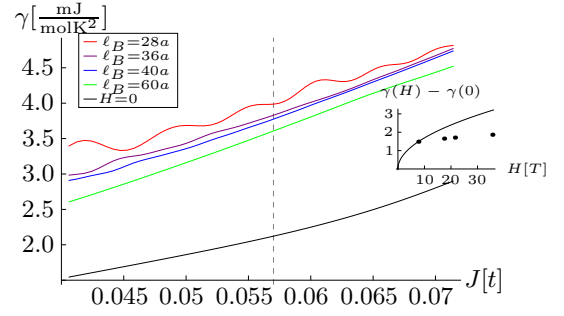}
\end{center}
 \caption{The Sommerfeld coefficient $\gamma(H,J)$ as a function of loop current order $J$, for four different finite fields with magnetic length $\ell_B=28a$ (red), $36a$ (purple), $40a$ (blue) and $60a$ (green) and zero field (black) at temperature 5K. In the inset, the four dots correspond to the intersection of the four finite field curves and the vertical dashed line in the main figure. For comparison, the solid line in the inset shows  $\sqrt{H}$ dependence; the units of the vertical axis are the same as in the main figure.}\label{fig:4H_5K}
\end{figure}

The experimental results show that the background on top of which
the oscillations reside has a $\sqrt{H}$ behavior\cite{Riggs}. In
Fig.\ref{fig:4H_5K} we show $\gamma(H,J)$ vs. $J$ for the four
finite fields and zero field at 5K. Although increasing with $H$,
the high temperature $\gamma(H,J)$ deviates from $\sqrt{H}$
behavior, as shown in the inset of Fig.\ref{fig:4H_5K}. As $J$ increases, the
finite field curves become closer to each other, but farther from
the zero field curve.

\subsubsection{Temperature dependence of the oscillations}
\begin{figure}[t]
\begin{center}
\includegraphics[width=0.4\textwidth]{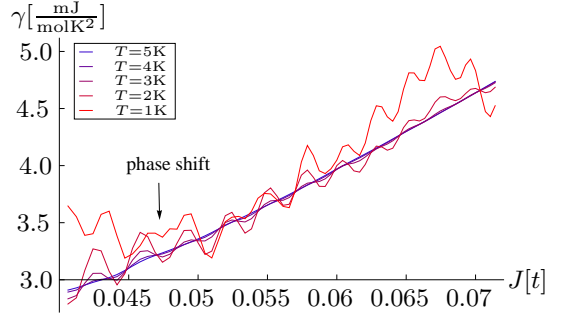}
\end{center}
 \caption{The Sommerfeld coefficient $\gamma(H,J)$ as a function of loop current order $J$ at temperature from 5K to 1K with step 1K, in the field with magnetic length $\ell_B=40a$. The phase shift basically agrees with that predicted by LK formula.}\label{fig:temp}
\end{figure}

The temperature dependence of the quantum oscillation of
$\gamma(H,J)$ is shown in Fig.\ref{fig:temp} for $\ell_B=40a\Rightarrow H\approx 17.4$T. At
$T\gtrsim 5$K, no oscillations appear. As the temperature is lowered, the
oscillations arise and the amplitude grows with decreasing
temperature. A phase shift is observed at 1K. In LK formula Eq.
(\ref{eq:LK}), a phase shift for $p=1$ occurs at $f''(x)=0$, where $x=2\pi^2
k_B T/(\hbar\omega_c)$. Here $\omega_c=eH/(m^*c)$. If we had charged Dirac particles, $m^*$ should be replaced by $E_F/(v_F v_\Delta)$. If the oscillations obey LK formula, then
with different parameter configurations, the phase shift should
occur at $x\approx 1.6$. Plugging in the parameters $T=1K$ and $H=17.4$T, using $E_F=J_0$ where $J_0$ is defined in Eq.(\ref{eq:J0}), and $e^*=0.7e$, we find that
the phase shift should be located at $J\approx0.05t$. This basically agrees with the numerical result in Fig.\ref{fig:temp}.

At small $J$, $x\sim JT/H$. Therefore, for a fixed magnetic field $H$, the phase shift should occur at the value of $J$
proportional to $1/T$. Similarly at different magnetic fields $H$, if
$T$ is fixed, the phase shift should occur at the value of $J$ proportional to
$H$. However, the phase shift can not be well identified in all the
cases, thus we are not able to accurately test whether this relation holds.

\subsubsection{Dependence of the oscillations on the configuration of the vortex lattice}
We also study the oscillations in a vortex lattice with the two vortices
placed at different positions in a magnetic unit
cell, while the size of the magnetic unit cell is kept the same. We
choose $\ell_B=40a$. The comparison is shown in Fig.\ref{fig:Vortex
position}. The blue curve is the result with vortices distributed
uniformly, with the separation $\ell_B/2=20a$ in both the horizontal
and the vertical directions(see Fig.\ref{fig:MUC}), the same as the
blue one in Fig.\ref{fig:4H_2K}; the red curve is the result with
the two vortices in the same magnetic unit cell placed much closer,
with the separation $4a$ in both directions. The same frequency of
oscillations is observed, excluding the possibility that the
oscillations are from Bragg plane reflections due to specific vortex
configurations. There is a difference between the phases of the two
configurations, which suggests that in the resistive state with
creeping vortices, the phase difference may smear out the oscillations.
\begin{figure}[t]
\begin{center}
\includegraphics[width=0.4\textwidth]{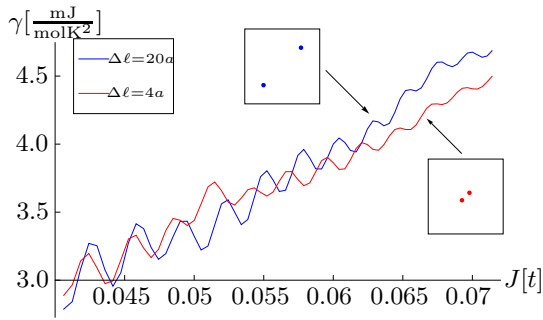}
\end{center}
 \caption{Oscillations of Sommerfeld coefficient, for different configurations of the vortex lattice. The separation between two vortices within a magnetic unit cell in both directions is $20a$ (blue) and $4a$ (red), in the field with $\ell_B=40a$.} \label{fig:Vortex position}
\end{figure}
\subsubsection{Density of states}
\begin{figure}[h]
\begin{center}
\includegraphics[width=0.5\textwidth]{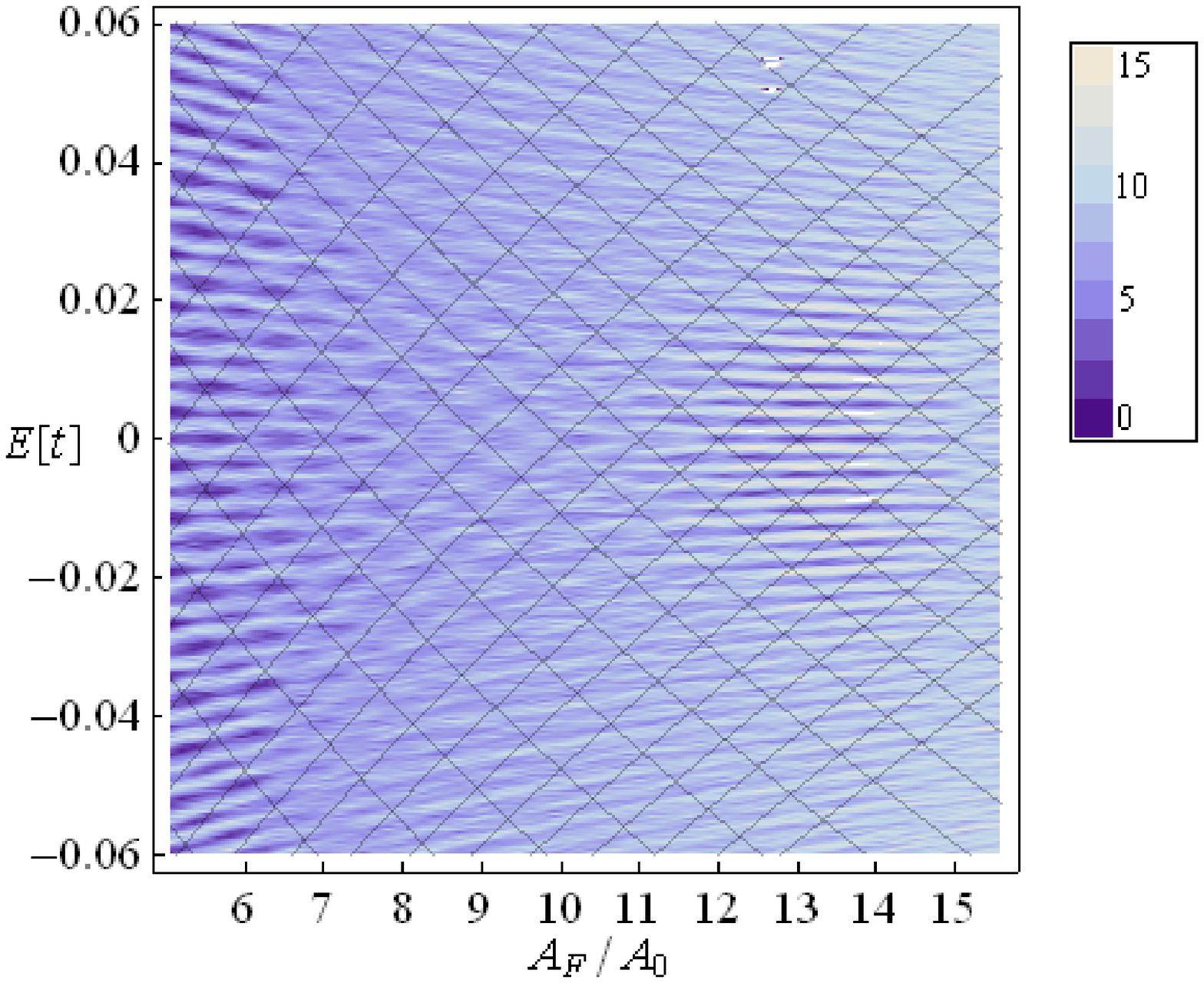}
\includegraphics[width=0.4\textwidth]{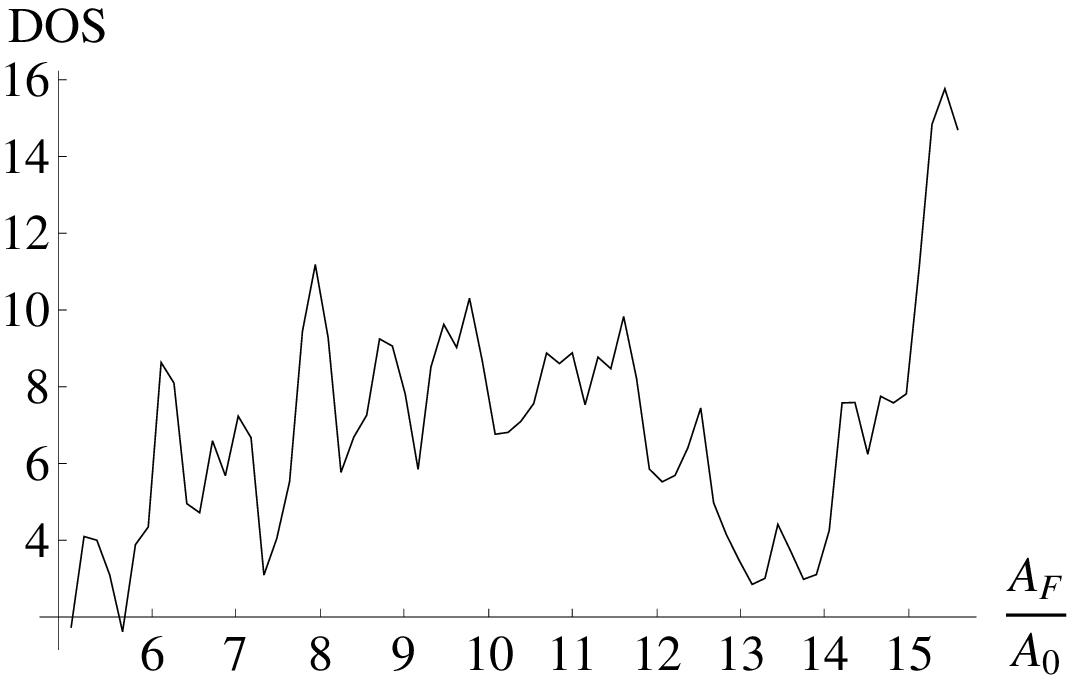}
\end{center}
 \caption{(Upper) The DOS as a function of energy and the Fermi pocket area, $A_F$, rescaled by $A_0=\frac{4\pi^2}{\ell^2_B}$. Here $\ell_B=40a$. The mesh shows the DOS of Landau quantized anisotropic Dirac fermions in the same field. The scale shown is such that the integrated DOS over the range of data is normalized to 1. (Lower) The DOS as a function of $A_F/A_0$ at zero energy. It does not oscillate smoothly, but at finite temperatures, states with nonzero energy are involved, making the oscillations in the specific heat smooth.} \label{fig:DOS}
\end{figure}

At zero temperature, the oscillations of the Sommerfeld coefficient simply corresponds to the oscillations of the density of states (DOS) at zero energy. We now investigate whether the DOS resembles that of Dirac fermions minimally coupled to an external magnetic field $H$. This can help us to illustrate the difference between the two systems. The DOS as a function of Fermi pocket area in the field $H$ with $\ell_B=40a$ is shown in Fig.\ref{fig:DOS}. The mesh shows the DOS of Dirac fermions with the same anisotropy and Fermi velocity minimally coupled to the same field $H$. Clearly, to a large extent, the $d$-wave superconductor with loop current order in the vortex state does not resemble Dirac fermions minimally coupled to a magnetic field. Therefore, the LK formula, which is derived for systems with Landau level quantization, does not hold in this case. Nevertheless, an oscillatory feature of the DOS is seen. Due to finite size effects, the zero energy DOS does not show smooth oscillations, but at intermediate temperatures, involvement of thermally activated modes with energy $\sim{\mathcal O}(k_BT)$ makes the oscillations of the specific heat smooth. It is this feature which is responsible for the oscillations of the specific heat at intermediate temperatures presented in Figs.(\ref{fig:4H_2K}-\ref{fig:Vortex position}).
\subsection{Oscillations as a function of Zeeman energy}
\begin{figure}[t]
\begin{center}
\includegraphics[width=0.4\textwidth]{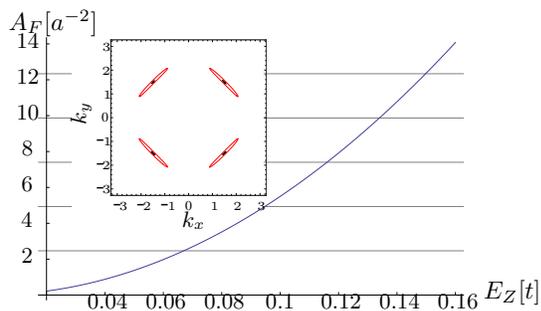}
\end{center}
 \caption{The dependence of Fermi pockets area $A_F$ on Zeeman energy in the $d$-wave superconducting state. (Inset) The four Fermi pockets at $E_Z=0.02t$ (black) and $E_Z=0.16t$ (red). The parameters are the same as in Fig.\ref{fig:FSarea}.} \label{fig:FS_Zeeman}
\end{figure}

\begin{figure}[t]
\begin{center}
\includegraphics[width=0.4\textwidth]{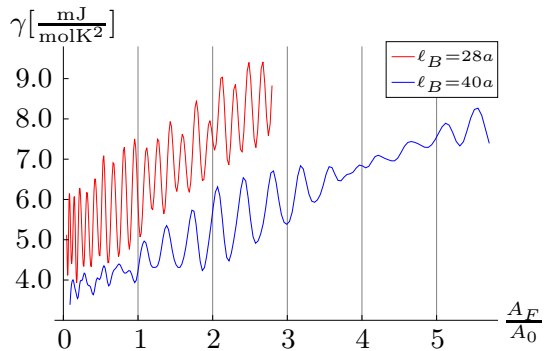}
\end{center}
 \caption{The Sommerfeld coefficient $\gamma(H)$ as a function of the rescaled area the Fermi pockets induced by Zeeman energy, in the magnetic fields with $\ell_B=40a$ (blue) and $\ell_B=28a$ (red).}\label{fig:Zeeman}
\end{figure}

Up to now, we have neglected the Zeeman splitting due to the external magnetic
field, since it only affects the results in an insignificant way and
the conclusions do not change. We imagine changing the Zeeman term
while holding the magnetic field and the pairing term fixed. This
will also induce Fermi pockets in the $d$-wave superconducting
state, even in the absence of the loop current order. In zero field, the
dispersion with such a term is
\begin{eqnarray}
E_\bk=\pm\sqrt{\xi_\bk^2+\Delta_\bk^2}+E_{Z}
\end{eqnarray}
where the Zeeman term $E_Z$ shifts the energies, resulting in four
Fermi pockets(see Fig.\ref{fig:FS_Zeeman}). Since at small energies
the pockets are ellipses, the area of each one $A_F$ is easily
calculated, and varies quadratically with Zeeman energy as shown in
Fig.\ref{fig:FS_Zeeman}. We use the same parameters as in the loop
current order case, and sweep $E_Z$ from $0.02t$ to $0.16t$, for two
magnetic fields with $\ell_B$=$28a$ and $\ell_B$=$40a$. In
Fig.\ref{fig:Zeeman} we show $\gamma(H)$ vs. $A_F/A_0$. If Landau
levels are formed and Onsager relation holds, the frequency of the
two oscillations should be the same, regardless of the magnitude of
the magnetic field. Nevertheless, the frequency is doubled when the
field is doubled, which is consistent with QP's forming Bloch
bands instead of Landau levels\cite{FT}. Comparing this result with
the oscillations induced by the loop current order, we conclude that the
latter has a more intricate nature whose effects can not be accounted merely by
the presence of the Fermi pockets. Rather, we believe, the special nature of the coupling between the loop
current order and the nodal $d_{x^2-y^2}$ QP's is essential to account for the observed oscillations obeying Onsager relation.

\section{The linearized problem}
\begin{figure}
\begin{center}
\includegraphics[width=0.4\textwidth]{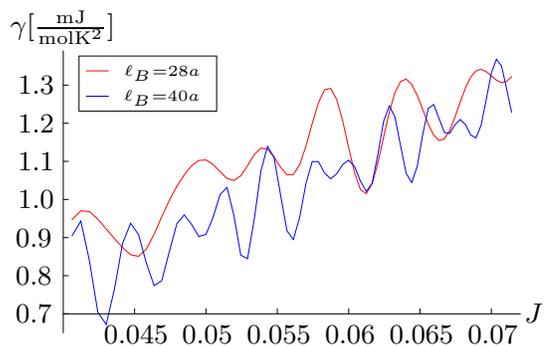}
\end{center}
 \caption{The Sommerfeld coefficient $\gamma(H,J)$ as a function of $J$ in fields with $\ell_B=28a$(red) and $\ell_B=40a$ (blue) at 5K resulting from one Fermi pocket in the linearized formulation.}
 \label{fig:40vs28linH}
\end{figure}

In the continuum formulation we diagonalize the linearized
Hamiltonian Eq.(\ref{eq:linH}) in the plane wave
basis\cite{FT,MelikyanTesanovic}, and repeat the calculations above
for two values of the magnetic field. The Fermi pocket area $A_F=\pi E_F^2/(v_F
v_\Delta)$, where $E_F=J_0$ given below Eq.(\ref{eq:J0}). We find that the two pockets give the same
contribution to oscillations of the specific heat, while the nodes
do not contribute to the oscillations. Fig.\ref{fig:40vs28linH}
shows the oscillations of $\gamma(H,J)$ in fields with $\ell_B=28a$
and $\ell_B=40a$ from one of the pockets. With the same range of $J$
as in Sec. \MakeUppercase{\romannumeral 3}, the same frequency is
found for both curves, which confirms our findings above. The
effective charge $e^*\approx e$ here.

\begin{figure}
\begin{center}
\includegraphics[width=0.4\textwidth]{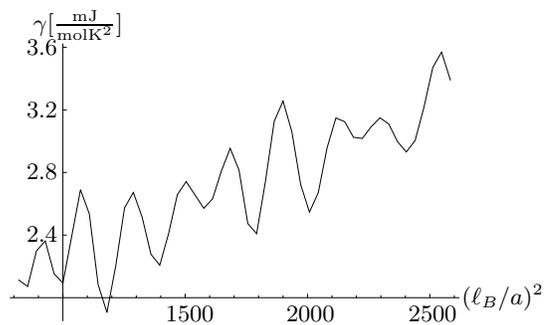}
\end{center}
 \caption{$\gamma(H,J)$ as a function of $\ell_B^2$ at $J=0.05t$ and $T=$2K, resulting from one Fermi pocket in the linearized formulation.}
 \label{fig:Cv_H}
\end{figure}

The linearized formulation enables us to sweep the external magnetic
field continuously at a fixed $J$, and to verify Onsager relation, which reads\cite{A&M}
\begin{eqnarray}\label{eq:OnsagerOther}
\Delta(\frac{1}{H})=\frac{2\pi e}{\hbar c A_F}
\end{eqnarray}
where $\Delta(1/H)$ is the period of oscillations if $A_F$ is fixed and $H$ is varied. This is equivalent to $\Delta(\ell_B^2)=(2\pi)^2/A_F$. We take $J=0.05t$ and sweep $\ell_B^2$, the result of which is shown in Fig.\ref{fig:Cv_H}. The period is $\Delta (\ell_B^2)\approx 200$. Using Eq.(\ref{eq:vel}), (\ref{eq:OnsagerOther}) and the expression of $J_0$, we determine that the effective charge $e^*\approx e$, which agrees with what is derived above for the tight-binding formulation.
\section{Discussions and Conclusions}
We have shown that the Fermi pockets induced by loop current order
can give rise to quantum oscillations of the specific heat within a
limited temperature range which seem to obey Onsager relation, with an
effective charge comparable to the electric charge.

Here we derive $J_0$ in two ways and compare them. We used the anisotropy $\al=14$ in our calculations, but the anisotropy of YB$_2$Cu$_3$O$_{6.56}$ studied in Ref.\cite{Riggs} is $\al\approx8$. To do the comparison, we assume that we would get quantum oscillations with the same effective charge for YB$_2$Cu$_3$O$_{6.56}$. Firstly, applying the standard formula of the specific heat in the absence of magnetic fields\cite{KivelsonVarma,Senthil_1}, we find $J_0\approx20\rm{meV}$ for $\gamma(0)\approx 2\rm{mJ/(mol\ K^2)}$. Secondly, using Onsager relation Eq.(\ref{eq:Onsager2}) and the frequency 531T\cite{Riggs}, we find $J_0\approx50\rm{meV}$ from quantum oscillations. They are of the same order, but differ by a factor of 2.5.

In the theory of
metals, Onsager relation is established from arguments that in a
uniform magnetic field, semiclassically, electrons move in constant
energy surfaces with quantized energies\cite{A&M}. In a $d$-wave
superconductor with loop current order, however, such a argument
does not work. Suppose that Bogoliubov QP's circulate the Fermi
surface which is an ellipse. Although the Fermi surface is still a
constant energy surface, the charge of a Bogoliubov QP varies with
its position as it moves around the Fermi surface, and the average
charge over the Fermi surface is zero\cite{KivelsonVarma}, which makes the
argument for metals invalid here.

We trace the origin of the oscillations observed in our numerical
calculation to the highly inhomogeneous fictitious magnetic field
experienced by the $d$-wave Dirac particles in the presence of the
loop order. Such coupling is absent if the Bogoliubov QP Fermi
pockets are due to Zeeman effect only. Indeed in the temperature
range where the oscillations appear, the Onsager relation holds
approximately in the case with the loop order but does not hold if
only the Zeeman shift is included.

One disadvantage of this picture is that the contribution from the
loop current order induced Fermi pockets is at odds with the
experimentally confirmed $\sqrt{H}$ behavior of the background on
top of which the oscillations occur. Another disadvantage is that the full Lifshitz-Kosevich relation
ceases to hold since ultimately there is no Landau quantization as shown in Fig.\ref{fig:DOS}.

Recently, a paper Ref.\cite{Senthil_1} with a similar concern
was published. Comparing with their work, we agree that there are
no quantized Landau levels near the Fermi pockets of Bogoliubov
QP's. Nevertheless, we do find quantum oscillations of the specific
heat in a small temperature range, while, by analyzing the density
of states at the Fermi level, they conclude that no quantum
oscillations should appear unless spin stripe order coexists with
loop current order.

\begin{acknowledgements}
We wish to thank S. A. Kivelson and C. M. Varma for useful discussions. This work was supported by a NSF CAREER award under Grant No. DMR-0955561, the NSF Cooperative Agreement No. DMR-0654118 and by the State of Florida .
\end{acknowledgements}
\appendix
\section{Quantum oscillations of the specific heat of Dirac fermions}
In this appendix we consider a hypothetical problem, the two-dimensional anisotropic Dirac fermions moving in a perpendicular magnetic field $H\hat{z}$. We use it to set up a contrast with the results found for the physical system considered in the main text. The Hamiltonian reads
\begin{eqnarray}
\mathcal{H}=v_F\left(\begin{array}{cc}0&\Pi_x-\frac{i}{\al_D}\Pi_y\\ \Pi_x+\frac{i}{\al_D}\Pi_y&0\end{array}\right)
\end{eqnarray}
where
\begin{eqnarray}
\vec{\Pi}=\vec{p}+\frac{e}{c}\vec{A}, \nabla\times\vec{A}=H\hat{z}
\end{eqnarray}
and $\al_D$ is the anisotropy. Since
\begin{eqnarray}
[\Pi_x,\Pi_y]=-i\hbar\frac{eH}{c}=-i\frac{\hbar^2}{\ell_B^2}
\end{eqnarray}
where $\ell_B=\sqrt{\frac{\hbar}{eH}}$ (defined differently from that in the main text), if we define the annihilation and creation operator as
\begin{eqnarray}
a&=&\frac{\sqrt{\al_D}\ell_B}{\sqrt{2}\hbar}(\Pi_x-\frac{i}{\al_D}\Pi_y),\\
a^\dagger&=&\frac{\sqrt{\al_D}\ell_B}{\sqrt{2}\hbar} (\Pi_x+\frac{i}{\al_D}\Pi_y),
\end{eqnarray}
then
\begin{eqnarray}
[a,a^\dagger]=1
\end{eqnarray}
and the Hamiltonian can be written as
\begin{eqnarray}
\mathcal{H}=\frac{\sqrt{2}\hbar v_F}{\sqrt{\al_D}\ell_B}\left(\begin{array}{cc}0&a\\a^\dagger&0\end{array}\right).
\end{eqnarray}
The eigenenergies of this Hamiltonian are
\begin{eqnarray}
\eps_n=\pm\sqrt{n}\frac{\sqrt{2}\hbar v_F}{\sqrt{\al_D}\ell_B}=\pm \zeta\sqrt{n}
\end{eqnarray}
where $\zeta=\sqrt{2}\frac{\hbar v_F}{\sqrt{\al_D}\ell_B}$. We consider only the positive eigenenergies $\eps_n=\zeta\sqrt{n}$ which are connected to the negative eigenenergies by particle-hole symmetry.

The grand potential is then
\begin{eqnarray}
\Omega&=&-TD\sum_{n=0}^\infty\ln(1+e^{-(\eps_n-\mu)/T})\\
&=&-TD\int_0^\infty d\eps\sum_{n=0}^\infty\delta(\eps-\eps_n)\ln(1+e^{-(\eps-\mu)/T})
\end{eqnarray}
where $D=gHL^2/\phi_0$ is the degeneracy of one Landau level, $g$ the number of species of fermions. Since
\begin{eqnarray}
\sum_{n=0}^\infty\delta(\eps-\sqrt{n}\zeta)=\sum_{n=0}^\infty\delta(n-\frac{\eps^2}{\zeta^2})\frac{2\eps}{\zeta^2},
\end{eqnarray}
then
\begin{eqnarray}
-\frac{\Omega}{TD}&=&\ln(1+e^{\frac{\mu}{T}})\nonumber\\
&+&\int_{0^+}^\infty d\eps\frac{2\eps}{\zeta^2}\sum_{n=0}^\infty\delta(n-\frac{\eps^2}{\zeta^2})\ln(1+e^{-\frac{\eps-\mu}{T}}).
\end{eqnarray}
Using
\begin{eqnarray}
\sum_{n=0}^\infty\delta(n-\frac{\eps^2}{\zeta^2})=\sum_{p=-\infty}^\infty e^{2\pi i p\frac{\eps^2}{\zeta^2}},
\end{eqnarray}
we have
\begin{eqnarray}
-\frac{\Omega}{TD}&=&\ln(1+e^{\frac{\mu}{T}})\nonumber\\
&+&\sum_{p=-\infty}^\infty\frac{1}{\zeta^2}\int_{0^+}^\infty d\eps 2\eps e^{2\pi ip\frac{\eps^2}{\zeta^2}}\ln(1+e^{-\frac{\eps-\mu}{T}}).
\end{eqnarray}
Integrating by parts, we have
\begin{widetext}
\begin{eqnarray}
-\frac{\Omega}{TD}
&=&\ln(1+e^{\frac{\mu}{T}})+\left.\ln(1+e^{-\frac{\eps-\mu}{T}})\frac{1}{2\pi ip}e^{2\pi ip\frac{\eps^2}{\zeta^2}}\right|_0^\infty+\int_0^\infty \frac{1}{2\pi ip}e^{2\pi ip\frac{\eps^2}{\zeta^2}}\frac{1}{T}\frac{1}{1+e^{\frac{\eps-\mu}{T}}}d\eps\nonumber\\
&=&\ln(1+e^{\frac{\mu}{T}})-\frac{1}{2\pi ip}\ln(1+e^{\mu/T})+\int_0^\infty \frac{1}{2\pi ip}e^{2\pi ip\frac{\eps^2}{\zeta^2}}\frac{1}{T}\frac{1}{1+e^{\frac{\eps-\mu}{T}}}d\eps.
\end{eqnarray}
\end{widetext}
The first two terms are non-oscillatory. Integrating by parts again for the third (oscillatory) term, we get
\begin{eqnarray}
-\frac{\Omega_{osc}}{TD}=\frac{1}{2\pi ip}\frac{1}{4T^2}\int_0^\infty d\eps\Phi(\eps)\frac{1}{\cosh^2\frac{\eps-\mu}{2T}}
\end{eqnarray}
where
\begin{eqnarray}
\Phi(\eps)=\int_0^\eps dye^{2\pi ip\frac{y^2}{\zeta^2}}.
\end{eqnarray}
Let $x=(\eps-\mu)/T$, then
\begin{eqnarray}
-\frac{\Omega_{osc}}{TD}&=&\frac{1}{8\pi ip}\frac{1}{T}\int_{-\frac{\mu}{T}}^\infty dx\frac{1}{\cosh^2\frac{x}{2}}\int_0^{Tx+\mu} dye^{2\pi ip\frac{y^2}{\zeta^2}}.\nonumber\\
\end{eqnarray}
Let $y=\mu\xi$, then
\begin{eqnarray}
-\frac{\Omega_{osc}}{\mu D}&=&\sum_{p\neq0}\frac{1}{8\pi ip}\int_{-\frac{\mu}{T}}^\infty \frac{dx}{\cosh^2\frac{x}{2}}\int_0^{1+\frac{T}{\mu}x} d\xi e^{2\pi ip\frac{\mu^2\xi^2}{\zeta^2}}.\nonumber\\
\end{eqnarray}
Using the formula
\begin{eqnarray}
\frac{\partial}{\partial z}\int^{y(z)}dxf(x)g(z,x)&=&\frac{\partial y(z)}{\partial z}f(y(z))g(z,y(z))\nonumber\\&+&\int^{y(z)}dxf(x)\frac{\partial g(z,x)}{\partial z},
\end{eqnarray}
and differentiating the grand potential twice with respect to $T$, we arrive at the oscillatory part of the Sommerfeld coefficient
\begin{widetext}
\begin{eqnarray}\label{eq:exact}
\frac{C_{osc}}{TL^2}&=&\frac{gH}{2\phi_0}\frac{\mu}{\zeta^2}\sum_{p\neq0}e^{2\pi ip\frac{\mu^2}{\zeta^2}}\int_{-\frac{\mu}{T}}^\infty dx\frac{x^2}{\cosh^2\frac{x}{2}}e^{4\pi ip\frac{\mu}{T}\frac{T^2}{\zeta^2}x}e^{2\pi ip\frac{T^2}{\zeta^2}x^2}(1+\frac{T}{\mu}x).
\end{eqnarray}
\end{widetext}
The first exponential accounts for the quantum oscillations of the Sommerfeld coefficient with the Fermi pocket area and the magnetic field, and the integral determines the amplitude as well as the phase shift of the oscillations.

\begin{figure}[t]
\begin{center}
\includegraphics[width=0.4\textwidth]{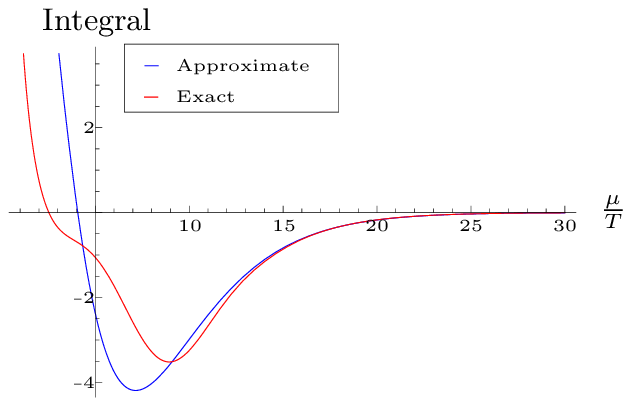}
\end{center}
 \caption{The integral appearing in Eq.(\ref{eq:approx})} (blue) and Eq.(\ref{eq:exact}) (red), with $T/\zeta=0.1$.
 \label{fig:append}
\end{figure}
At the low-temperature limit, $\mu/T\gg 1$, we can extend the lower limit of the first integral to $-\infty$ since the difference is exponentially small. Also, in the integral, the first exponential oscillates much faster than the second one since $x$ is bounded by $1/\cosh^2\frac{x}{2}$ to $\sim1$. Thus we drop the second exponential as well as the last factor,
\begin{eqnarray}\label{eq:approx}
\frac{C_{osc}}{TL^2}&=&\frac{gH}{2\phi_0}\frac{\mu}{\zeta^2}\sum_{p\neq0}e^{2\pi ip\frac{\mu^2}{\zeta^2}}\int_{-\infty}^\infty dx\frac{x^2}{\cosh^2\frac{x}{2}}e^{4\pi ip\frac{T\mu}{\zeta^2}x}\nonumber\\
\end{eqnarray}
The integral has the analytical form
\begin{eqnarray}
\int_{-\infty}^\infty dx\frac{x^2}{\cosh^2\frac{x}{2}}e^{i\frac{\lambda}{\pi} x}&=&-4\pi^2f''(\lambda)\nonumber\\
\end{eqnarray}
where
\begin{eqnarray}
f(\lambda)=\frac{\lambda}{\sinh\lambda}.
\end{eqnarray}
Now,
\begin{eqnarray}
\frac{C_{osc}}{TL^2}&=&-4\pi^2\frac{gH}{2\phi_0}\frac{\mu}{\zeta^2}\sum_{p\neq0}e^{2\pi ip\frac{\mu^2}{\zeta^2}}f''(z)\\ &=&-4\pi^2\frac{gH}{\phi_0}\frac{\mu}{\zeta^2}\sum_{p=1}^\infty \cos\left(2\pi p\frac{\mu^2}{\zeta^2}\right)f''(z)
\end{eqnarray}
where
\begin{eqnarray}
z=4\pi^2p\frac{T\mu}{\zeta^2}.
\end{eqnarray}
The Fermi pocket area is
\begin{eqnarray}
A_F=\frac{\al_D\pi \mu^2}{\hbar^2v_F^2},
\end{eqnarray}
and the cyclotron "mass" is
\begin{eqnarray}
m^*=\left.\frac{\hbar^2}{2\pi}\frac{\partial A}{\partial E}\right|_{E=\mu}=\frac{\al_D\mu}{v_F^2}.
\end{eqnarray}
So the combinations appearing in the Sommerfeld coefficient are
\begin{eqnarray}
\frac{\mu}{\zeta^2}&=&\frac{\al_D\mu}{2v_F^2\hbar eH}=\frac{m^*}{2\hbar eH}=\frac{1}{2\hbar\omega_c},\\
\frac{\mu^2}{\zeta^2}&=&\frac{\al_D\mu^2}{2v_F^2\hbar eH}=\frac{A_F h}{4\pi^2eH}=\frac{A_F\phi_0}{4\pi^2H},
\end{eqnarray}
thus
\begin{eqnarray}
\frac{C_{osc}}{TL^2}&=&-\pi\frac{gm^*}{\hbar^2}\sum_{p=1}^\infty \cos\left(p\frac{A_F\phi_0}{2\pi H}\right)f''(z)
\end{eqnarray}
with
\begin{eqnarray}
z=2\pi^2p\frac{T}{\hbar\omega_c}.
\end{eqnarray}
The coefficient is exactly the same as in the Schr\"{o}dinger case, except that $m^*$ has a different definition and depends on $\mu$ now. In the Schr\"{o}dinger case, $C_{osc}/T$ is strictly periodic in the Fermi pocket area since the amplitude has no dependence on $\mu$, but for Dirac fermions it is different. Because $z$ depends on $\mu$, the amplitude, as well as the phase shift, depend not only on the temperature and the magnetic field, but also on the Fermi pocket area.

From the first harmonic $p=1$, we can deduce the Onsager relation: the period of the cosine as a function of $A_F$ and $1/H$, respectively, is
\begin{eqnarray}
\Delta(A_F)=\frac{4\pi^2H}{\phi_0},\label{eq:Onsager1}\\
\Delta(\frac{1}{H})=\frac{4\pi^2}{\phi_0 A_F}.\label{eq:Onsager2}
\end{eqnarray}
To compare the approximate expression of the amplitude obtained in the limit $\mu/T\gg1$ with the expression valid to any $\mu/T$, in Fig.\ref{fig:append} we plot the integral in Eq.(\ref{eq:approx}) and that in Eq.(\ref{eq:exact}) as a function of $\mu/T$, while fixing $T/\zeta=0.1$. The two curves basically coincide at $\mu/T\gtrsim 10$. The position where the phase shift occurs, i.e. where the integral changes sign, is different for the two curves. At a smaller $T/\zeta$, the phase shift occurs at a larger $\mu/T$ for both curves, and the positions where the phase shift occurs become closer.

\bibliography{osc}
\end{document}